# Optimal Orientations of Quartz Crystals for Bulk Acoustic Wave Resonators with the Consideration of Thermal Properties

J. Wang[a,*], L. M. Zhang[a], S.Y. Wang[a], L. T. Xie[a], B. Huang[a], T. F. Ma[a], J. K. Du[a], M. C. Chao[b],

S. Shen[b], R. X. Wu[c], H. F. Zhang[d]

[1]Piezoelectric Device Laboratory, Ningbo University, Ningbo, Zhejiang 315211, China

[2]TXC (Ningbo) Corporation, 189 Huangshan West Road, Beilun, Ningbo, Zhejiang 315800, China

[3]Department of Civil Engineering, Ningbo Polytechnic, 1069 Xinda Road, Belun, Ningbo, Zhejiang 315800, China

[4]Department of Mechanical and Energy Engineering, University of North Texas, Denton, TX 76210, USA

*Corresponding author.  Email: wangji@nbu.edu.cn; Tel.: 0574-87600303.

Piezoelectric crystals are widely used for acoustic wave resonators of different functioning modes and types including BAW and SAW. It is well-known that only some special orientations of crystals will exhibit desirable properties such as mode couplings, thermal sensitivity, acceleration sensitivity, and others that are important in design and applications of resonators. With extensive studies on physical properties in last decades and increasing industrial needs of novel products, it is necessary to comb the known knowledge of quartz crystal material for novel orientations and better products as agendas in the industry. With known material properties like elastic, piezoelectric, dielectric, and thermal constants, we can establish the relationships between vibrations and bias fields such as temperature to ensure a resonator immunizing from excessive response to changes causing significant degradation of resonator properties and performances. Since the theoretical framework of wave propagation in piezoelectric solids is known, we need to use the existing data and results for the validation of current orientations in actual products. The agreement will give us needed confidence of the theory and analytical procedures. Through rotations, we calculated physical properties as functions of angles and bias fields, enabling the calculation of resonator properties for the identification of optimal cuts. Such a procedure can also be applied to similar crystals for a careful examination of possible orientations to maximize the potential use of materials in acoustic wave resonators

**Keywords:** Vibration; Resonator; Quartz; Temperature; Frequency; Orientation

## 1. INTRODUCTION

The study on the relationship between the frequency and temperature of a resonator has a long history. Shortly after the resonator was first invented in the 1920s, Koga set the first-order derivative of the frequency-temperature relationship to zero for a single-rotation crystal plate at 25°C, inventing AT- and BT-cut crystal resonators.[1-3] In 1962, Bechmann solved the equation of the first derivative of the frequency to temperature of the doubly-rotated crystal at 25°C systematically and obtained the curves of the two angles of first-order zero temperature coefficients of the thickness-shear vibrations.[4] In 1978, EerNisse considered the effect of stress on the basis of temperature-frequency results of Bechmann and obtained a doubly-rotated SC-cut with stress compensation and excellent temperature effect.[5] Based on good temperature-frequency behavior, both AT- and SC-cut resonators are the most widely used and stable products.[6-7] However, the finding of these two cuts is from the first derivative of the temperature-frequency relationship. In practice, what is really needed for more stable cuts is a crystal resonator which the first and second derivatives of the frequency-temperature relationship are zeroes at the same time. Existing research is still limited to solving the first derivative of the equation as zero. Therefore, it is necessary for us to solve the system of equations with the first- and second-order derivatives as zeroes.

Using the basic equations of motion in incremental thermal fields by Lee[8-9] and then degenerating them into equations of infinite plates[10-11], we can calculate the frequency by solving the eigenvalue problem. Elastic constants and coefficients of thermal expansion are obtained by coordinating transformation from those values in a standard coordinate system. In this way we can establish the function of frequency with respect to temperature and two angles. Then setting the first- and second-order derivative to zeroes with respect to the temperature, we obtain the first- and second-order zero temperature coefficients curve with two angles and temperature in three dimensions, which are responding to crystal cut and reference temperature respectively.

## 2. THEORY

### A. BASIC EQUATIONS OF QUARTZ CRYSTAL IN A THERMAL FIELD

Temperature affects crystal vibrations in two ways. First, the free object will expand when the temperature rises. While the object is bounded by boundaries, the thermal stress will be generated. Second, the elastic constants change as the temperature changes. Bechmann first used the basic equations of elasticity in Eulerian description to obtain the Christoffel's equation with the assumption of straight crested waves. The elastic constants in this equation take the temperature effect into consideration, and then take the expanded density and boundary conditions into account to obtain the frequency. However, in 1986, Lee established the basic equation which is based on the Lagrangian description, the density, thickness, and orientation of the crystal plate are chosen to be constant values in their natural state. In addition, quartz is a crystal with weak piezoelectric effect, thus we neglect the effect of the piezoelectricity in this study.

The strains with thermal consideration are

$$e_{ij} = \frac{1}{2}(\beta_{ki}u_{k,j} + \beta_{kj}u_{k,i}), \qquad (1)$$

where $u_i$, $e_{ij}$, $\beta_{ij}$ are the components of incremental displacement, strain and thermal expansion coefficients, respectively, and

$$\beta_{ik} = \delta_{ij} + \alpha_{ij}^{(1)}(T - T_0) + \alpha_{ij}^{(2)}(T - T_0)^2 + \alpha_{ij}^{(3)}(T - T_0)^3, \qquad (2)$$

where $\delta_{ij}$ is Kronecker symbol, $T$ is the actual temperature, $T_0$ is the reference temperature which is 25°C, $\alpha_{ij}^{(k)}, (k = 1, 2, 3)$ is the $k$th-order expansion coefficient, and the values are chosen from Lee[8-9].

The thermal stress-strain relations are

$$t_{ij} = D_{ijkl}e_{kl}, \qquad (3)$$

where $t_{ij}$ and $D_{ijkl}$ are stress and thermal elastic constants, and

$$D_{ijkl} = C_{ijkl} + D_{ijkl}^{(1)}(T - T_0) + D_{ijkl}^{(2)}(T - T_0)^2 + D_{ijkl}^{(3)}(T - T_0)^3, \qquad (4)$$

where $C_{ijkl}$ and $D_{ijkl}^{(m)}, (m = 1, 2, 3)$ are the elastic constants, the $m$th-order elastic constants coefficients of quartz crystal, and the values are chosen from Lee[8-9].

The stress equations of motion are

$$\beta_{ik}t_{jk,j} = \rho \ddot{u}_i \quad \text{in } V, \qquad (5)$$

and the traction-stress relations in boundary are

$$p_i = n_j(\beta_{ik}t_{jk}) \quad \text{in } S, \qquad (6)$$

where $t_{ij}, p_i, n_j$ are the stress tensor, traction vector and the surface normal base vector.

### B. EQUATIONS OF AN INFINITE PLATE IN AN INCREMENTAL THERMAL FIELD

Quartz resonators usually work in thickness-shear mode. Because the model is simple, the calculation of frequency of the infinite plate is straight-forward. Previous studies and experiments show that the results of the thickness-shear frequency and the temperature effect are useful references in product development. The infinite plate vibration equation without temperature effect has been studied extensively.[2,10,11]

For infinite plates, the length and width coordinates are ignored and only the thickness coordinates are left. In this case, the displacement function can be written as

$$u_i = u_i(x_2, t), \qquad (7)$$

and the boundary conditions of the plate can be expressed as
$$T_2 = T_4 = T_6 = 0, \quad x_2 = \pm b. \tag{8}$$
Now we rewrite the tensor form of Eqs. (1) - (6) in matrix form as:
gradient equations
$$\begin{bmatrix} S_6 \\ S_2 \\ S_4 \end{bmatrix} = \begin{bmatrix} \beta_{11} & \beta_{12} & \beta_{13} \\ \beta_{12} & \beta_{22} & \beta_{23} \\ \beta_{13} & \beta_{23} & \beta_{33} \end{bmatrix} \begin{bmatrix} u_{1,2} \\ u_{2,2} \\ u_{3,2} \end{bmatrix}, \tag{9}$$
constitutive equations
$$\begin{bmatrix} T_6 \\ T_2 \\ T_4 \end{bmatrix} = \begin{bmatrix} D_{66} & D_{26} & D_{46} \\ D_{26} & D_{22} & D_{24} \\ D_{46} & D_{24} & D_{44} \end{bmatrix} \begin{bmatrix} S_6 \\ S_2 \\ S_4 \end{bmatrix}, \tag{10}$$
equations of motion
$$\rho \begin{bmatrix} \ddot{u}_1 \\ \ddot{u}_2 \\ \ddot{u}_3 \end{bmatrix} = \begin{bmatrix} \beta_{11} & \beta_{12} & \beta_{13} \\ \beta_{12} & \beta_{22} & \beta_{23} \\ \beta_{13} & \beta_{23} & \beta_{33} \end{bmatrix} \begin{bmatrix} T_{6,2} \\ T_{2,2} \\ T_{4,2} \end{bmatrix}, \tag{11}$$
by substituting Eq. (9) into the Eq. (10), and then into Eq. (11) resulting the following matrix equation
$$\rho \begin{bmatrix} \ddot{u}_1 \\ \ddot{u}_2 \\ \ddot{u}_3 \end{bmatrix} = K \begin{bmatrix} u_{1,2} \\ u_{2,2} \\ u_{3,2} \end{bmatrix}, \tag{12}$$
where
$$K = \begin{bmatrix} \beta_{11} & \beta_{12} & \beta_{13} \\ \beta_{12} & \beta_{22} & \beta_{23} \\ \beta_{13} & \beta_{23} & \beta_{33} \end{bmatrix} \begin{bmatrix} D_{66} & D_{26} & D_{46} \\ D_{26} & D_{22} & D_{24} \\ D_{46} & D_{24} & D_{44} \end{bmatrix} \begin{bmatrix} \beta_{11} & \beta_{12} & \beta_{13} \\ \beta_{12} & \beta_{22} & \beta_{23} \\ \beta_{13} & \beta_{23} & \beta_{33} \end{bmatrix}. \tag{13}$$

Solving above equation, we first assume a special solution $u_i = A_i \sin(\eta x_2) e^{i\omega t}$, and substitute the solution into Eq. (12) to obtain the eigenvalue equation. Using the principle of superposition of linear equations, we can get the general solution of the system, defining wave speed as $c = \rho \omega^2 / \eta^2$, we can get the following eigenvalue equation
$$KA = cA. \tag{14}$$
Solving the above eigenvalue equation, according to the principle of superposition, the displacement solution can be expressed as $u_i = \sum_{j=1}^{3} A_{ij} \sin(\eta_2 x_2)$. By substituting it into the boundary conditions, we can get
$$\begin{bmatrix} T_6 \\ T_2 \\ T_4 \end{bmatrix} = \begin{bmatrix} D_{66} & D_{26} & D_{46} \\ D_{26} & D_{22} & D_{24} \\ D_{46} & D_{24} & D_{44} \end{bmatrix} \begin{bmatrix} \beta_{11} & \beta_{12} & \beta_{13} \\ \beta_{12} & \beta_{22} & \beta_{23} \\ \beta_{13} & \beta_{23} & \beta_{33} \end{bmatrix} \begin{bmatrix} \sin \eta_1 b & \sin \eta_2 b & \sin \eta_3 b \\ \sin \eta_1 b & \sin \eta_2 b & \sin \eta_3 b \\ \sin \eta_1 b & \sin \eta_2 b & \sin \eta_3 b \end{bmatrix} \begin{bmatrix} A_{31} \\ A_{32} \\ A_{33} \end{bmatrix} = 0. \tag{15}$$
Equation (15) has a nontrivial solution if and only if the determinant of the matrix is zero, or
$$\sin \eta_1 b = 0 \text{ or } \sin \eta_2 b = 0 \text{ or } \sin \eta_2 b = 0, \tag{16}$$
now the wavenumber is
$$\eta_i = \frac{n\pi}{b}, i = 1,2,3, \tag{17}$$
and the angular frequency
$$\omega_i = \frac{n\pi}{b} \sqrt{\frac{c_i}{\rho}}, i = 1,2,3, \tag{18}$$
finally the frequency

$$f_i = \frac{n}{2b}\sqrt{\frac{c_i}{\rho}}, \quad i = 1,2,3. \tag{19}$$

In the above we choose $f_1 > f_2 > f_3$, and their vibration modes are designated as A, B, and C mode, respectively.

## C. DERIVATIVE EQUATIONS

In fact, the resonator shows small frequency change over the operating temperature range, which can be described by letting both the first- and second-order derivatives of the temperature-frequency function be zero. Because the frequency is always positive, in order to facilitate the calculation, we write the above equation in logarithmic form,[1]

$$\ln f = \ln\frac{n}{2} - \ln b + \frac{1}{2}\ln c - \frac{1}{2}\ln\rho. \tag{20}$$

Because the thickness and density are chosen from the natural state, they are constants at 25°C. The wave speed changes with temperature so that its derivative and the second derivative are zero, resulting in the following two equations

$$T_f^{(1)}(\varphi,\theta,T) = \frac{d\ln f}{dT} = \frac{1}{2c}\frac{dc}{dT} = 0, \tag{21}$$

$$T_f^{(2)}(\varphi,\theta,T) = \frac{dT_f^{(1)}}{dT} = -\frac{1}{2}\left(\frac{1}{c}\frac{dc}{dT}\right)^2 + \frac{1}{2c}\frac{d^2c}{dT^2} = 0. \tag{22}$$

Solving the above two equations simultaneously, we get a curve with two angles and temperature as three independent variables.

## 3. RESULTS

The quartz belongs to the 32 symmetry group, according to Bechmann's research, we set the rotation angles in the range of $\varphi \in (0°, 30°)$ and $\theta \in (-90°, 90°)$.[4] Bechmann first systematically studied the first-order zero temperature coefficient point of the C-mode of infinite plates at 25°C as a function of angle[4], as shown in Fig. 1. For comparison, we superimposed the first- and second-order zero temperature coefficients curve projected in plane of two angles to the Fig. 1. The three-dimensional curve is shown in Figs. 2-4 which are in a large temperature range of $(-200°C, 350°C)$.

From Figs. 1-4, we get the following conclusions:

- The first-order zero temperature coefficients curve and the positive branch of first- and second-order zero temperature coefficients curve are very close in the range $\varphi \in (0°, 15°)$, so the AT-cut resonator using the properties of the first derivative is equal to zero whose second derivative is also equal to zero. In addition, as the angle increases, it will drift downward. This also explains why the accurate angle $\theta = 33.93°$ of SC-cut is smaller than the angle $\theta = 34.3°$ solved by the equations of zero stress and first-order derivative of frequency temperature.[5]
- The first- and second-order zero temperature coefficients curves are very different from the first-order zero temperature coefficients curve in $\theta < 0$. It has two branches, decreasing with the angle of $\varphi$. We can see that BT-cut does not fall on the first- and second-order zero temperature curve, so its temperature-frequency curve is parabolic, which is inferior to AT- and SC-cut frequency temperature curve as we know.
- From Figs. 2-4 we can see that the branch of $\theta > 0$, the first- and second-order zero temperature point will also gradually increase as $\varphi$, which explains why the temperature of a SC-cut the temperature of zero frequency is 95°C instead of 25°C. When $\varphi > 25°$ the

- first- and second-order zero temperature point increases dramatically to 300°C, with the potential for making high temperature resonators.
- At $\theta < 0$ the first- and second-order zero temperature coefficient point is below $-60°C$, with the potential to make low-temperature resonators.

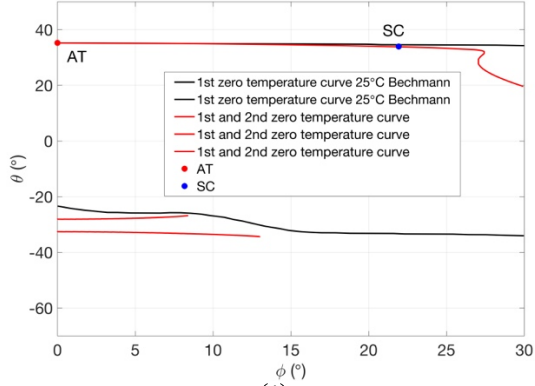

Figure 1. Locus of $T_f^{(1)} = 0$ by Bechmann and $T_f^{(1)} = 0$, $T_f^{(2)} = 0$ for mode C.

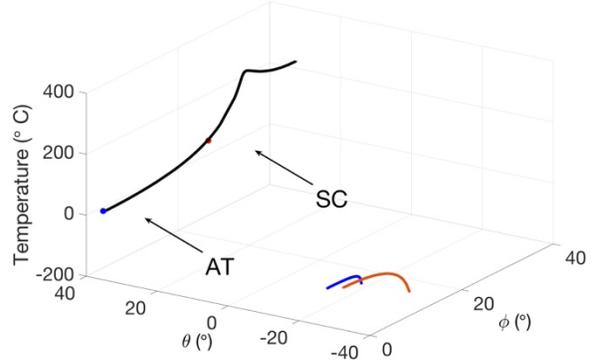

Figure 2. Three-dimensional curves of $T_f^{(1)} = 0$ by Bechmann and $T_f^{(1)} = 0$, $T_f^{(2)} = 0$ for mode C.

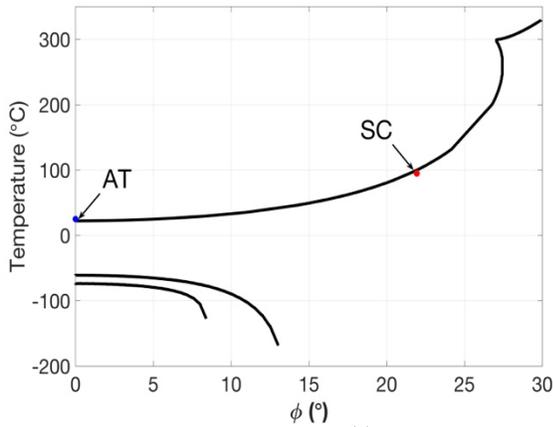

Figure 3. Projection curves $T_f^{(1)} = 0$ by Bechmann and $T_f^{(1)} = 0$, $T_f^{(2)} = 0$ for mode C.

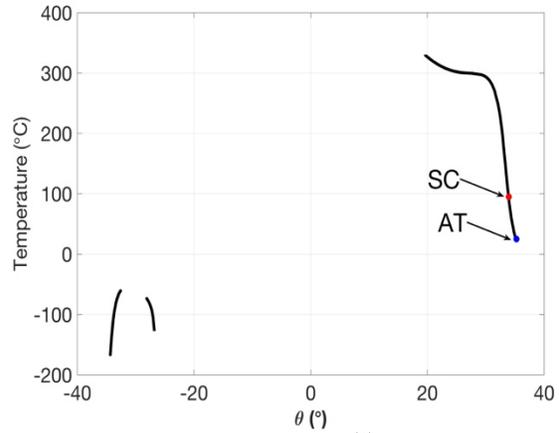

Figure 4. Projection curves $T_f^{(1)} = 0$ by Bechmann and $T_f^{(1)} = 0$, $T_f^{(2)} = 0$ for mode C.

In order to clearly show that the points on the first- and second-order zero temperature curves do have excellent frequency-temperature performance, we select some points evenly on the top three curves to draw their temperature-frequency curves, as shown in Figs. 5-8.

## 4. CONCLUSIONS AND DISCUSSIONS

Using the first- and second-order derivatives of the temperature-frequency equations, we have solved the curve of the first- and second-order zero temperature coefficients with the arbitrary angle of doubly-rotated crystal. On these curves, resonators have the same excellent temperature-frequency characteristics as the most commonly used AT- and SC-cut types. This is a general procedure to find good temperature-frequency effect, which can provide reference for the future research on the relationship between frequency and temperature of other crystal materials. We compared the first-order zero temperature coefficient curve that Bechmann solved at 25°C and found that the positive branch $\varphi \in (0°, 25°)$ are close to each other. In the past, the AT- and SC-cut resonators obtained through the first derivative of the temperature-frequency relationship have the property that the second derivative is also zero. In addition, we found that two branches of $\varphi \in (25°, 30°)$ and $\theta < 0$ have very high and low second-order zero temperature coefficient points, implying the potential to create high- and low-temperature resonators.

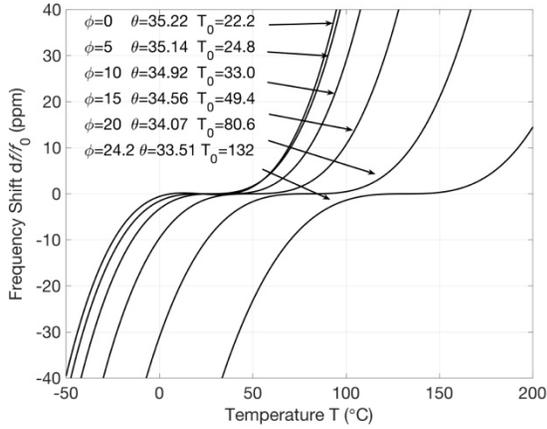

Figure 5. Frequency-temperature curves of different angles of positive branch in $\varphi \in (0°, 25°)$.

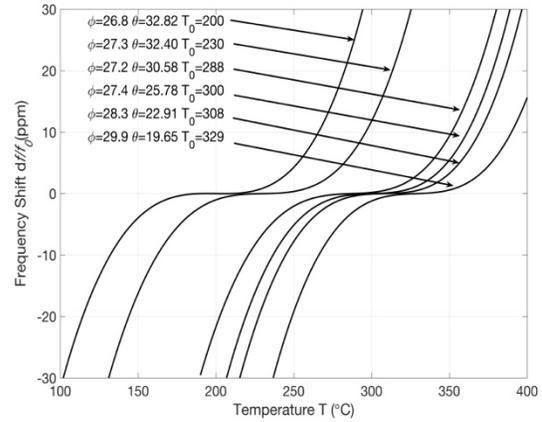

Figure 6. Frequency-temperature curves of different angles of positive branch in $\varphi \in (25°, 30°)$.

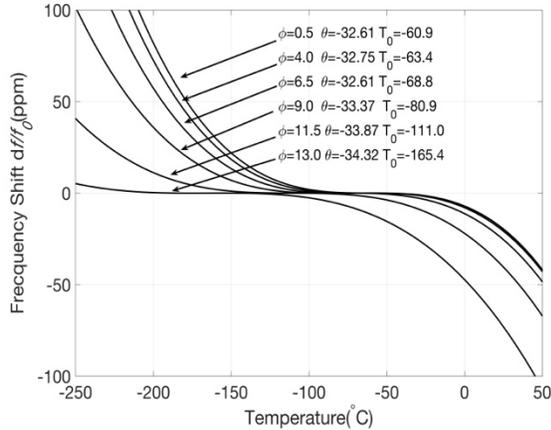

Figure 7. Frequency-temperature curves of different angles of middle branch in Fig.4.

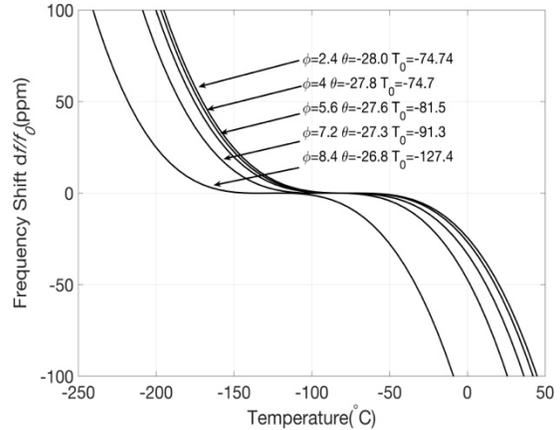

Figure 8. Frequency-temperature curves of different angles of left branch in Fig.4.

## ACKNOWLEDGMENTS

This research is supported in part by the National Natural Science Foundation of China (Grants 11372145, 11672142, & 11772163). Additional support is from the TXC (Ningbo) Corporation under industrial partnership program with Ningbo University. The research is also supported by the K. C. Wong Magana Fund established and administered by Ningbo University.